# LAWS OF CONSERVATION OF MOMENTUM AND ANGULAR MOMENTUM IN CLASSICAL ELECTRODYNAMICS OF MATERIAL MEDIA

Alexander L Kholmetskii[1], Oleg V. Missevitch[2] and T Yarman[3,4]

[1]Belarusian State University, Minsk, Belarus, e-mail: khol123@yahoo.com
[2]Institute for Nuclear Problems, Belarusian State University, Minsk, Belarus
[3]Okan University, Akfirat, Istanbul, Turkey
[4]Savronik, Eskisehir, Turkey

We analyze the laws of conservation of momentum and angular momentum in classical electrodynamics of material media with bound charges, and explore the possibility to describe the properties of such media via a discrete set of point-like charges of zero size (as imposed by special relativity), and via continuous charge/current distributions. This way we put a question: do we have to recognize the infinite fields at the location of elementary charges as the essential physical requirement, or such infinite fields can be ignored via introduction of continuous charge distribution? In order to answer this question, we consider the interaction of a homogeneously charged insulating plate with a compact magnetic dipole, moving along the plate. We arrive at the apparent violation of the angular momentum conservation law and show that this law is recovered, when the electric field at the location of each elementary charge of the plate is taken infinite. This result signifies that the description of electromagnetic properties of material media via the continuous charge and current distributions is not a universal approximation, and at the fundamental level, we have to deal with a system of elementary discrete charges of zero size, at least in the analysis of laws of conservation of momentum and angular momentum.

## 1. Introduction

In the present contribution we address to the known inconsistency of CED, related to the relativistic requirement of zero size of an elementary charge, when the proper fields of this charge become infinite at its location [1, 2]. As is known, this leads to the infinite electromagnetic (EM) mass of the classical electron, and a similar problem of infinite proper energy of electron emerges in quantum electrodynamics, too (see, e.g. [3]). Such an infinite energy is eliminated via the procedure of standard renormalization [1, 4], though its correctness from the mathematical viewpoint remains questionable (see, e.g. [1, 5]). In these conditions, even a purely classical approach to further elaboration of the problem of infinite EM energy of elementary charge remains topical. This way a crucial question emerges: do we have to recognize the infinite EM fields in a vicinity of elementary charge as a physical reality, or such infinite fields can be ignored via the adoption of finite size of elementary charge and/or introduction of continuous charge distribution? The goal of the present contribution is to get answer on this question, analyzing the conservation laws of total momentum and total angular momentum for an isolated system of interacting classical charges bound in material media.

More specifically, we will analyze the time variation of mechanical and *interactional* electromagnetic momentum and angular momentum, ignoring the proper EM momentum of each individual charge, as is usually done in the analysis of energy-momentum conservation law in classical electrodynamics [1, 4, 6]. By such a way, for example, we exclude the known "4/3 problem" for the classical electron [7], though a radical solution of this problem has been recently suggested in ref. [8].

Within the framework of this restriction, we first address to the paper by Page and Adams [9], where the authors demonstrated that for two interacting charged particles, a misbalance of their mutual forces is exactly counteracted by the time derivative of momentum of interactional EM field of these particles, while a misbalance of torque is counteracted by variation of the an-gular momentum of interactional EM field. Here we would like to notice that in the determination of field momentum and field angular momentum, the authors of ref. [9] carried out integration of corresponding densities over the *entire* space and thus, they tacitly adopted *zero* size of elementary charge. We mention that other authors of similar numerous publications (not cited here), generalizing the results of ref. [9] to an arbitrary number of interacting charged particles, also carrying out the integration of field momentum density and field angular momentum density over the entire space and thus, again tacitly adopted zero size of the classical electron. Therefore, at the location of each elementary charge, its electric and magnetic fields become infinite, though such field singularities occur removable in the determination of *interactional* field momentum and field angular momentum.

At the same time, now we point out that classical electrodynamics in material media usually implies the description of media via continuous charge and current distributions, which, in general, is at odds with the relativistic requirement of point-like elementary charges with infinite fields at their location. It is obvious that the dissonance between the representation about a set of discrete charges and about continuous charge distributions can be much softened, if we adopt a finite size of elementary charge, avoiding by such a way any singularities of electric/magnetic fields. However, in this case the fundamental problem about the interaction of a system of charged particles [9] should be again reanalyzed with a separate integration of field momentum density and field angular momentum density over the *free* space, and over the inner volume of each charged particle; and we could not state *a priori* that the laws of conservation of total momentum and total angular momentum would be fulfilled. Besides, the result of such integration depends on the adopted charge distribution inside the elementary charge.

In this respect, it is much more convenient to verify the laws of conservation of total momentum and total angular momentum, dealing with charged media, characterized by continuous charge and current distributions.

We explore this problem in section 2 with a particular example: the interaction of a moving magnetic dipole with a resting insulating homogeneously charged plate of large size and demonstrate the fulfillment of total momentum conservation law, but, at the same time, the apparent violation of total angular momentum conservation law. In section 3 we discuss the results obtained and show that the implementation of the law of conservation of total momentum occurs insensitive to the hypothesis about a size of elementary charge and thus, it is fulfilled in both alternative descriptions of charged media: via a set of discrete charges or via a continuous charge and current distributions. However, it is not the case for the law of conservation of total angular momentum, which can be fulfilled only via the adoption of infinite EM field at the location of each elementary charge. Thus, at the fundamental level we have to take zero size of such elementary charges, and the introduction of continuous charge distribution in material media represents, in general, the incorrect approximation, when the law of conservation of total angular momentum is analyzed.

## 2. Magnetic dipole moving along a charged plate.

We consider a homogeneously charged insulating plate with the surface charge density $\sigma$ and an electrically neutral compact magnetic dipole, moving along this plate (see Fig. 1). The width of the plate is equal to $2h$, and it lies in the plane $yz$, so that its electric field $\boldsymbol{E}$ is parallel to the axis $x$ near the plate in the spatial regions far from its boundaries in $yz$-plane. We further assume that the sizes of the plate along the axis $y$ ($Y_0$) and axis $z$ ($Z_0$) are so large, that the field of magnetic dipole becomes negligible on its edges. The dipole is moving along the axis $y$ at the velocity $v$, and is separated from the plate by the distance $x_0 \ll Y_0, Z_0$. This configuration is sometimes considered with respect to the Aharonov-Casher effect [10], though we suppose that at $t=0$, the proper magnetic dipole moment of dipole $\boldsymbol{m}_0$ is parallel to the electric field $\boldsymbol{E}$, and this effect is vanishing. We want to determine the mutual forces between the plate and magnetic dipole and corresponding torques, as well as to calculate a time variation of momentum and angular mo-mentum of interactional EM field, in order to verify the implementation of the total momentum and total angular momentum conservation laws.

In further analysis, we use the following expression for the force on a compact dipole, moving at velocity $v$ in an electric $E$ and magnetic $B$ fields

$$F = \nabla(p \cdot E) + \nabla(m \cdot B) + \frac{1}{c}\frac{d}{dt}(p \times B) - \frac{1}{c}\frac{d}{dt}(m \times E), \qquad (1)$$

which is derived on the basis of Lorentz force law via the approach suggested by Vekstein [11] with further inclusion of hidden momentum contribution [12]. Here the electric $p$ and magnetic $m$ dipole moments are related to the proper vectors $p_0$, $m_0$ by the known relationships [13]

$$p = p_0 - \frac{(\gamma-1)}{\gamma v^2}(p_0 \cdot v)v + \frac{v \times m_0}{c}, \quad m = m_0 - \frac{(\gamma-1)}{\gamma v^2}(m_0 \cdot v)v + \frac{p_0 \times v}{c} \qquad (2a\text{-}b)$$

where $\gamma = (1-v^2/c^2)^{-1/2}$ is the Lorentz factor.

The corresponding expression for the torque exerted on the dipole has the form [14]

$$T = p \times E + m \times B + \frac{1}{c}v \times (p \times B) - \frac{1}{c}v \times (m \times E). \qquad (3)$$

Applying eqs. (7-9) to the configuration in Fig. 1, where at the initial time moment the magnetic dipole moment $m$ is parallel to $E$, we first observe that due to the constancy of $E$ at the location of magnetic dipole, the Coulomb force on the electric dipole moment (2a) (the first term on rhs of eq. (1)) is equal to zero. The corresponding torque component reads as

$$T_{mC} = p \times E = \frac{1}{c}((v \times m_0) \times E) = -\frac{1}{c}v(E \cdot m_0) + \frac{1}{c}m_0(E \cdot v) = -\frac{1}{c}v(E \cdot m_0). \qquad (4)$$

where we have used the vector identity

$$a \times (b \times c) = b(a \cdot c) - c(a \cdot b), \qquad (5)$$

and taken into account the equality $E \cdot v = 0$ for the configuration of Fig. 1.

Due to the torque (4), the magnetic dipole begins to rotate in the plane $xz$, and the time derivative of its magnetic dipole moment is defined by the equation

$$dm/dt = \omega \times m, \qquad (6)$$

where we designated $\omega\{0, -\omega, 0\}$ the angular rotation frequency at the time moments near $t=0$.

Using eq. (6), we further evaluate the force component on magnetic dipole due to time variation of hidden momentum (the fourth term on rhs of eq. (1)), which takes the form

$$F_{mh} = -\frac{1}{c}\frac{d}{dt}(m \times E) = -\frac{1}{c}((\omega \times m) \times E) = \frac{1}{c}\omega(E \cdot m_0), \qquad (7)$$

where we have used the vector identity (6) and taken into account that $\omega \cdot E = 0$.

Addressing now to eq. (3), we notice that the rotation of magnetic dipole in the plane $xz$ does not charge the fourth term on its right-hand side (hidden momentum contribution), which remains equal to zero and thus, eqs. (4) and (7) define the total torque and the total force on the dipole, correspondingly.

Now let us calculate the force and torque on the charged plate due to magnetic dipole, which can be induced by the two components of electric field of magnetic dipole: Coulomb field component of electric dipole (2a) and inductive field component $E_I = -\partial A/c\partial t$ generated by magnetic dipole at the location of the plate.

It is obvious that the Coulomb force on the plate due to electric dipole (2a) is equal to zero, while the related torque $T_{PC}$ is equal to the torque on the dipole (4) with the reverse sign:

$$T_{PC} = -T_{mC} = \frac{1}{c}v(E \cdot m_0) \qquad (8)$$

In order to evaluate the inductive electric field of magnetic dipole, we will use for simplicity a weak relativistic limit, corresponding to the accuracy of calculations $c^{-2}$, which is sufficient for the purposes of the present paper. In this approximation, we can put the Lorentz factor $\gamma=1$ in the relativistic transformation of magnetic/electric dipole (eqs. (2)) and fields/potentials (see, e.g. [6]), as well as to ignore the scale contraction effect, because the expressions for the force and torque for the configuration of Fig. 1 already contain the factor $c^{-1}$. Thus, in the adopted approximation we can use the non-relativistic expression for the vector potential of magnetic dipole [6] $\boldsymbol{A} = \boldsymbol{m}_0 \times \boldsymbol{R}_m / R_m^3$ in the rest frame of plate, where the vector $\boldsymbol{R}_m$ has the components $\{-x_0, y, z\}$ on the plate. Hence,

$$\boldsymbol{E}_I = -\frac{1}{c}\frac{\partial \boldsymbol{A}}{\partial t} = -\frac{1}{c}\frac{(\boldsymbol{\omega} \times \boldsymbol{m}_0) \times \boldsymbol{R}_m}{R_m^3} - \frac{1}{c}\frac{\partial}{\partial t}\frac{\boldsymbol{m}_0 \times \boldsymbol{R}_m}{R_m^3}\bigg|_{\boldsymbol{m}_0 = \text{constant}} \quad (9)$$

The first term on rhs describes the inductive electric field component due to rotation of magnetic dipole moment $\boldsymbol{m}$ (eq. (6)), whereas the second term corresponds to the inductive field component due to the displacement of dipole along the axis $y$ at fixed spatial orientation of $\boldsymbol{m}_0$.

First we evaluate the second term on rhs of eq. (9), designating it as $\boldsymbol{E}_I^{motion}$ and taking into account that at fixed $\boldsymbol{m}_0$, this term can be presented in the form

$$\boldsymbol{E}_I^{motion} = -\frac{1}{c}\frac{\partial \boldsymbol{A}^{motion}}{\partial t} = -\frac{1}{c}\frac{\partial}{\partial t}\frac{\boldsymbol{m}_0 \times \boldsymbol{R}_m}{R_m^3}\bigg|_{\boldsymbol{m}_0 = \text{constant}} = -\frac{v}{c}\frac{\partial}{\partial y}\frac{\boldsymbol{m}_0 \times \boldsymbol{R}_m}{R_m^3}. \quad (10)$$

Next we calculate the first term on rhs of eq. (9), designating it as

$$\boldsymbol{E}_I^{rotation}{}_I = -\frac{1}{c}\frac{\partial \boldsymbol{A}^{rotation}}{\partial t} = -\frac{1}{cR_m^3}((\boldsymbol{\omega} \times \boldsymbol{m}_0) \times \boldsymbol{R}_m) = \frac{1}{cR_m^3}\boldsymbol{\omega}(\boldsymbol{R}_m \cdot \boldsymbol{m}_0) - \frac{1}{cR_m^3}\boldsymbol{m}_0(\boldsymbol{R}_m \cdot \boldsymbol{\omega}). \quad (11)$$

The corresponding force and torque on the plate are determined by the equations

$$\boldsymbol{F}_{PI} = \int_{-Y_0}^{Y_0}\int_{-Z_0}^{Z_0} \boldsymbol{E}_I \sigma dz dy, \quad \boldsymbol{T}_{PI} = \int_{-Y_0}^{Y_0}\int_{-Z_0}^{Z_0} (\boldsymbol{r}_{yz} \times \boldsymbol{E}_I)\sigma dz dy, \quad (12\text{a-b})$$

where $\boldsymbol{r}_{yz}\{0, y, z\}$ is the radius-vector lying along the plate.

Substituting eq. (10) into eqs. (12a-b), we find that both the force and torque on the plate due to inductive field component $\boldsymbol{E}_I^{motion}$ are equal to zero at $Y_0, Z_0 \gg x_0$. Inserting eq. (11) into eq. (12a), we obtain the force on the plate due to inductive electric field of dipole:

$$\boldsymbol{F}_{PI} = -\boldsymbol{\omega}\frac{x_0 m_0 \sigma}{c}\int_{-Y_0}^{Y_0}\int_{-Z_0}^{Z_0}\frac{dzdy}{(x_0^2 + y^2 + z^2)^{3/2}} - \boldsymbol{m}_0\frac{\omega\sigma}{c}\int_{-Y_0}^{Y_0}\int_{-Z_0}^{Z_0}\frac{ydzdy}{(x_0^2 + y^2 + z^2)^{3/2}} = -\frac{1}{c}\boldsymbol{\omega}(\boldsymbol{E}\cdot\boldsymbol{m}_0), \quad (13)$$

where we have used the expression

$$\int\frac{dx}{(a^2 + x^2)^{3/2}} = \frac{x}{a^2(a^2 + x^2)^{1/2}} + const,$$

and taken into account the equality $E = 2\pi\sigma$ for homogeneously charged plate.

We see that in the adopted accuracy of calculations (where we can put $\boldsymbol{m} = \boldsymbol{m}_0$), the force (7) on the magnetic dipole is equal with the reverse sign to the force on the plate (13). One can show that the balance of these forces is kept in the exact relativistic analysis, too. It means that for the configuration in Fig. 1,

$$d\boldsymbol{P}_M/dt = 0, \quad (14)$$

where $\boldsymbol{P}_M$ stands for the mechanical momentum of the system.

Next, let us show that the time derivative of hidden momentum of magnetic dipole $\boldsymbol{P}_h = (\boldsymbol{m}_0 \times \boldsymbol{E})/c$ is equal to the time derivative of field momentum with the reverse sign, according to the general proof, given in refs. [15, 16] for a magnetic dipole in a static electric field.

The time derivative of field momentum for the configuration of Fig. 1 is equal to

$$d\boldsymbol{P}_{EM}/dt = \frac{1}{4\pi c}\int_V (\boldsymbol{E}\times\dot{\boldsymbol{B}})dV, \quad (15)$$

where we designated $\dot{\boldsymbol{B}} = d\boldsymbol{B}/dt$, and $V$ stands for the entire space. For the problem of Fig. 1, the latter adoption is legitimate in the limit $h \to 0$ (very thin plate), if we assume that the electric field inside the plate has *finite* values at its any inner point.

Further we notice that the equality $E=const$, taking place outside the plate in the spatial regions of a non-vanishing vector potential of magnetic dipole, cannot be extended to infinity in any real situation. Thus, we carry out further integration over a finite spatial volume $V_0$, being restricted by the coordinates $(-X_0…-h, +h…X_0)$ along the axis $x$, $(-Y_0… Y_0)$ along the axis $y$, and $(-Z_0… Z_0)$ along the axis $z$, and adopt the constancy of $E$ in this region. Due to the condition $X_0, Y_0, Z_0 >> x_0$, the magnetic field of dipole practically disappears outside the volume $V_0$, so that the result of integration of eq. (15) over $V$ and $V_0$ is practically the same. Further, we designate $E_x=-E$ (at $-X_0<x<-h$), and $E_x=+E$ (at $X_0>x>h$) within the volume $V_0$. Then, for any smooth function $f(x, y, z)$, we obtain in the limit $h \to 0$

$$\int_{-X_0}^{X_0} E_x f(x,y,z) dx = -E \int_{-X_0}^{0} f(x,y,z) dx + E \int_{0}^{X_0} f(x,y,z) dx. \tag{16}$$

Replacing in eq. (15) the volume $V$ by $V_0$, one can easily see that at $X_0, Y_0, Z_0 >> x_0$, the $x$- and $z$-components of time derivative of field momentum (15) are equal to zero, and we have a single non-vanishing $y$-component

$$d(P_{EM})_y / dt = -\frac{1}{4\pi c} \int_{V_0} E_x \dot{B}_z dV = -\frac{1}{4\pi c} \int_{V_0} E_x \left( \frac{\partial \dot{A}_y}{\partial x} - \frac{\partial \dot{A}_x}{\partial y} \right) dxdydz, \tag{17}$$

where we have used the equality $\boldsymbol{B} = \nabla \times \boldsymbol{A}$.

Evaluating the integral (17), we address to eqs. (10), (11) and notice that the time variation of component of vector potential $\boldsymbol{A}^{motion}$ does not contribute to the time variation of field momentum. This result is obvious, because the motion of magnetic dipole along the plate far from its edges does not change the field momentum and field angular momentum, when the magnetic dipole moment $\boldsymbol{m}_0$ is fixed. Hence, only the component $\boldsymbol{A}^{rotation}$ is relevant, and hereinafter we designate its time derivative as

$$\dot{\boldsymbol{A}} = \frac{1}{r_m^3} \boldsymbol{m}_0 (\boldsymbol{r}_m \cdot \boldsymbol{\omega}) - \frac{1}{r_m^3} \boldsymbol{\omega}(\boldsymbol{r}_m \cdot \boldsymbol{m}_0)$$

(compare with eq. (11)), where we replaced the vector $\boldsymbol{R}_m$ {$-x_0$, $y$, $z$} by vector $\boldsymbol{r}_m$ {$(x-x_0)$, $y$, $z$} for an arbitrary point $x$. Thus, the vector $\dot{\boldsymbol{A}}$ has the vanishing $z$-component, and its other components read as

$$\dot{A}_x = -\frac{m_0 y \omega}{\left((x-x_0)^2 + y^2 + z^2\right)^{3/2}}, \quad \dot{A}_y = \frac{m_0 (x-x_0) \omega}{\left((x-x_0)^2 + y^2 + z^2\right)^{3/2}}, \tag{18a-b}$$

where we have taken into account that vector $\boldsymbol{\omega}$ has the components $\{0, -\omega, 0\}$.

The explicit calculation of the total time derivative of field momentum is presented in Appendix A (eq. (A7)), which can be written in the vector form as

$$\frac{d\boldsymbol{P}_{EM}}{dt} = \int_{entire\ space} \frac{d\boldsymbol{p}_{EM}}{dt} dV = \frac{1}{c} \boldsymbol{\omega} (\boldsymbol{E} \cdot \boldsymbol{m}_0), \tag{19}$$

which is equal to the force on magnetic dipole due to time variation of its hidden momentum (7).

Thus, taking into account that $d\boldsymbol{P}_h/dt = -\boldsymbol{F}_{mh} = -\frac{1}{c}\boldsymbol{\omega}(\boldsymbol{E}\cdot\boldsymbol{m}_0)$, we derive the conservation of total momentum for the isolated configuration in Fig. 1 (see eqs. (7), (14) and (19)):

$$\frac{d}{dt}(\boldsymbol{P}_M + \boldsymbol{P}_h + \boldsymbol{P}_{EM}) = 0. \tag{20}$$

In the light of a recent discussion [17-21] about the Einstein-Laub force law [22, 23]

$$\boldsymbol{F} = (\boldsymbol{P} \cdot \nabla)\boldsymbol{E} + (\boldsymbol{M} \cdot \nabla)\boldsymbol{B} + \frac{1}{c}\frac{\partial \boldsymbol{P}}{\partial t} \times \boldsymbol{B} - \frac{1}{c}\frac{\partial \boldsymbol{M}}{\partial t} \times \boldsymbol{E} \tag{21}$$

(where $\boldsymbol{P}$ is polarization, $\boldsymbol{M}$ magnetization, and the free charges are absent), the force component (7) is derived with the last term of eq. (21), and, like in the conventional Lorentz force law in material media, this term should be associated with the time variation of hidden momentum contribution. Otherwise, eq. (21) cannot be fulfilled.

Next we verify the law of conservation of total angular momentum for the problem in Fig. 1. For this purpose we further calculate the torque on the plate due to the inductive electric field (9) of magnetic dipole, as well as the time derivative of field angular momentum at the time moments near $t=0$.

This torque component on the plate is described by the equation

$$T_{PI} = \int_{-Y_0}^{Y_0}\int_{-Z_0}^{Z_0} (r_{yz} \times E_I^{motion})\sigma dzdy + \int_{-Y_0}^{Y_0}\int_{-Z_0}^{Z_0} (r_{yz} \times E_I^{rotation})\sigma dzdy . \quad (22)$$

We omit straightforward calculations, which prove that the first integral on rhs of eq. (22) is equal to zero. It means that the progressive motion of magnetic dipole along the plate at fixed $m_0$ does not induce a torque on the plate.

Next we evaluate the second integral in eq. (22), using eq. (11):

$$T_{PI}^{rotation} = \int_{-Y_0}^{Y_0}\int_{-Z_0}^{Z_0}\left(\frac{r_{yz} \times \omega(R_m \cdot m_0)}{cR_m^3}\right)\sigma dzdy - \int_{-Y_0}^{Y_0}\int_{-Z_0}^{Z_0}\left(\frac{r_{yz} \times m_0(R_m \cdot \omega)}{cR_m^3}\right)\sigma dzdy . \quad (23)$$

At the time moments near $t=0$, eq. (23) gives the vanishing $x$- and $y$-components of torque, whereas for the $z$-component we have:

$$\begin{aligned}(T_{PI}^{rotation})_z &= -\frac{m_0\omega\sigma}{c}\int_{-Y_0}^{Y_0}\int_{-Z_0}^{Z_0}\frac{y^2}{(x_0^2+y^2+z^2)^{3/2}}dzdy = -\frac{2m_0\omega\sigma Z_0}{c}\int_{-Y_0}^{Y_0}\frac{y^2}{(x_0^2+y^2)\sqrt{x_0^2+Z_0^2+y^2}}dy = \\ &= -\frac{2m_0\omega\sigma Z_0}{c}\left(\ln\frac{\sqrt{x_0^2+Z_0^2+Y_0^2}+Y_0}{\sqrt{x_0^2+Z_0^2+Y_0^2}-Y_0} - \frac{x_0\pi}{Z_0}\right) = -\frac{Em_0\omega}{\pi c}Z_0\ln\frac{\sqrt{2}+1}{\sqrt{2}-1} + \frac{Em_0\omega x_0}{c}\end{aligned} \quad (24)$$

at $Y_0=Z_0 \gg x_0$. (Here we have used eq. (A2), as well as the expression

$$\int\frac{y^2 dy}{(a^2+y^2)\sqrt{b^2+y^2}} = \ln(\sqrt{b^2+y^2}+y) - \frac{a}{\sqrt{b^2-a^2}}\arctan\frac{y\sqrt{b^2-a^2}}{a\sqrt{b^2+y^2}} + constant ).$$

The torque on the plate due to inductive electric field of dipole can be presented in the vector form

$$T_{PI} = \frac{x_0 \times \omega(E \cdot m_0)}{\pi c}\frac{Y_0}{x_0}\ln\frac{\sqrt{2}+1}{\sqrt{2}-1} - \frac{x_0 \times \omega(E \cdot m_0)}{c}, \quad (25)$$

where $x_0\{x_0, 0, 0\}$, $\omega\{0, -\omega, 0\}$ and $E=\{2\pi\sigma, 0, 0\}$.

The mechanical torque of the system consists of three components: torque (4) on the magnetic dipole, torques (8), (25) on the charged plate, as well as the torque constituted by the lever of forces (7), (13) acting on the dipole and plate, correspondingly, at their spatial separation $x_0$, i.e. $T_L = \frac{1}{c}x_0 \times \omega(E \cdot m_0)$ The time derivative of mechanical angular momentum $\theta_M$ of the system is defined by the sum of the mentioned torque components:

$$\frac{d\theta_M}{dt} = T_{mC} + T_{PC} + T_{PI} + T_L = \frac{x_0 \times \omega(E \cdot m_0)}{\pi c}\frac{Y_0}{x_0}\ln\frac{\sqrt{2}+1}{\sqrt{2}-1} \quad (26)$$

Finally, in order to check the conservation of total angular momentum, we calculate the time derivative of field angular momentum ($d\theta_{EM}/dt$)

$$\frac{d\theta_{EM}}{dt} = \frac{1}{4\pi c}\int_V r \times (E \times \dot{B})dV . \quad (27)$$

In order to save place, we omit straightforward calculations on the basis of eqs. (16) and (18a-b), which yield
$d(\theta_{EM})_x/dt = d(\theta_{EM})_y/dt = 0$,
and determine the non-vanishing $z$-component of field angular momentum:

$$\frac{(d\theta_{EM})_z}{dt} = -\frac{1}{4\pi c}\int_{V_0} xE_x\dot{B}_z dV = -\frac{1}{4\pi c}\int_{V_0} xE_x\left(\frac{\partial \dot{A}_y}{\partial x} - \frac{\partial \dot{A}_x}{\partial y}\right)dV, \qquad (28)$$

where we have taken into account that in the spatial region $V_0$ the electric field has a single non-vanishing component $E_x$, and also used the equality $\boldsymbol{B} = \nabla \times \boldsymbol{A}$.

The explicit calculation of $(d\theta_{EM})_z/dt$ is given in the Appendix B (eq. (B7)), which can be presented in the vector form as

$$\frac{d\boldsymbol{\theta}_{EM}}{dt} = -\frac{\boldsymbol{x}_0 \times \boldsymbol{\omega}(\boldsymbol{E}\cdot\boldsymbol{m}_0)}{\pi c}\frac{Y_0}{x_0}\ln\frac{\sqrt{2}+1}{\sqrt{2}-1} + \frac{\boldsymbol{x}_0 \times \boldsymbol{\omega}(\boldsymbol{E}\cdot\boldsymbol{m}_0)}{c}. \qquad (29)$$

Thus, we see that the sum of eqs. (26) and (29) yields the non-conservation of total angular momentum for the configuration of Fig. 1:

$$\frac{d\boldsymbol{\theta}_{EM}}{dt} + \frac{d\boldsymbol{\theta}_M}{dt} = \frac{\boldsymbol{x}_0 \times \boldsymbol{\omega}(\boldsymbol{E}\cdot\boldsymbol{m}_0)}{c} \ne 0. \qquad (30)$$

## 3. Discussion

Summarizing our findings for the problem in Fig. 1, we conclude:
- the forces of action and reaction between the magnetic dipole and charged plate exactly balance each other, and the sum of hidden momentum of magnetic dipole and field momentum represents the conserved quantity, which provide the implementation of total momentum conservation law (eq. (20);
- at time moments near $t=0$, the sum of time derivative of mechanical angular momentum in the system (eq. (26)), and the time derivative of angular momentum of EM field (eq. (29)) is not equal to zero, so that we face with the paradoxical situation, i.e. the violation of the law of conservation of total angular momentum for the problem in Fig. 1, expressed by eq. (30).

Seeking a way for the resolution of this paradox, we first pay attention on eq. (16), where, in fact, we ignored the contribution to the total momentum and angular momentum of EM fields inside the plate. It is a warranted approximation in the limit $h \to 0$, if we adopt the *finite* EM fields in each inner point of the plate. However, if the size of elementary charge is equal to zero, the electric field at the location of such charges situated inside the plate becomes infinite and thus, even in the limit $h \to 0$, we can get, in general, a finite contribution to the field momentum and angular momentum for the configuration of Fig. 1. Therefore, we have to evaluate one more contribution to the field momentum

$$(\boldsymbol{P}_{EM})_{in} = \frac{1}{4\pi c}\int_{-Z_0}^{Z_0}\int_{-Y_0}^{Y_0}\int_{-h}^{h}(\boldsymbol{E}_{in}\times\boldsymbol{B})dxdydz \qquad (31)$$

and field angular momentum

$$\frac{d(\boldsymbol{\theta}_{EM})_{in}}{dt} = \int_{-Z_0}^{Z_0}\int_{-Y_0}^{Y_0}\int_{-h}^{h}\boldsymbol{r}\times(\boldsymbol{E}_{in}\times\boldsymbol{B})dxdydz \qquad (32)$$

for the configuration of Fig. 1, even if the limit $h \to 0$ is implied. Here $\boldsymbol{E}_{in}$ stands for the inner electric field inside the plate.

Considering the integrals (31) and (32), we notice that in a vicinity of each elementary charge, the magnetic field of magnetic dipole $\boldsymbol{B}$ can be taken as the constant value. Then, adopting the electric field of each elementary charge $\boldsymbol{E}(\boldsymbol{r})$ to be radially symmetric in its own frame (as is usually adopted in CED [2]), we obtain the corresponding radial symmetry of the vector product $\boldsymbol{E}\times\boldsymbol{B}$ at constant $\boldsymbol{B}$. Therefore, the integration of this product over a small volume enclosed the elementary charge gives zero contribution to the field momentum even at the infinite electric field at the location of point-like elementary charge. Addressing to the problem in Fig. 1 and evaluating the integral (31), we can express this general result in other terms, observing that for any elementary charge $e$, belonging to the plate and lying in the plane $yz$, its location point $\{0, y_e, z_e\}$ represents the point of discontinuity of the $x$-component of the proper electric field $E_x$, where it changes the sign without changing its value. Since the magnetic field of dipole can be taken constant at $-h<x<h$ and fixed $y_e$, $z_e$, it is clear that the y-component of integral (31), containing $E_x$ (see eq. (17)) is averaged to zero. Hence, we can derive the conservation of total momentum (20), operating only with external fields. Therefore, no paradoxes emerge, when we calculate the total field momentum for the configuration of Fig. 1, operating only with EM fields outside material media with the negligible spatial volume.

However, evaluating now the integral (32), we stress that the double vector product $\mathbf{r}\times(\mathbf{E}\times\mathbf{B})$, in general, is not a radially symmetric function in the frame of each elementary charge, even for a radially symmetric electric field $\mathbf{E}(\mathbf{r})$ and at the adopted constancy of $\mathbf{B}$ near each charge. Therefore, the integration of the product $\mathbf{r}\times(\mathbf{E}\times\mathbf{B})$ over infinitesimal volume enclosing an elementary charge, in general, does not yield zero, when the proper field of this charge become infinite at its location point. Addressing again to the problem in Fig. 1 and evaluating the integral (32), we observe that for any elementary charge $e$, belonging to the plate and lying in the plane $yz$, the product $xE_x$ (which emerges in the calculation of z-component of field angular momentum, eq. (28)), has the same sign at $x>0$ and $x>0$ and fixed $y$, $z$. Therefore, the integral (32), in general, is not equal to zero at $h\to 0$, and $E\to\infty$, and its ignorance explains the apparent violation of the law of conservation of total angular momentum (30).

Thus, the adoption of infinite electric field at the location of elementary classical charges represents the necessary condition to recover the field angular momentum conservation law for the problem considered above. This statement is equivalent to the recognition of zero size of the classical electron, which, in general, makes inapplicable the approximation of continuous charge distribution inside the charged plate. Therefore, in the exact calculation of time derivative of field angular momentum contribution (27), we should deal with a set of discrete charges and their proper electric fields, being infinite at their location points. This problem, however, lies beyond the classical physics, and in the framework of CED we can only conjecture a particular expression for the electric field $\mathbf{E}_{in}$ in integral (32), which, being added to the rhs of eq. (30), provides the conservation of total angular momentum. Omitting particular calculations, we mention that at the choice

$$(E_{in})_x = 2\pi\sigma x_0 \frac{\delta(x)}{x}, \tag{33}$$

the integral (33) yields in the limit $h\to 0$:

$$\frac{d(\boldsymbol{\theta}_{EM})_{in}}{dt} = \int_{-Z_0}^{Z_0}\int_{-Y_0}^{Y_0}\int_{-h}^{h} \mathbf{r}\times(\mathbf{E}_{in}\times\mathbf{B})dxdydz = -\frac{\mathbf{x}_0\times\boldsymbol{\omega}(\mathbf{E}\cdot\mathbf{m}_0)}{c}.$$

Thus, adding $d(\boldsymbol{\theta}_{EM})_{in}/dt$ to lhs of eq. (30), we obtain

$$\frac{d\boldsymbol{\theta}_{EM}}{dt} + \frac{(d\boldsymbol{\theta}_{EM})_{in}}{dt} + \frac{d\boldsymbol{\theta}_M}{dt} = 0,$$

which expresses the conservation of total angular momentum for the problem in Fig 1.

At the same time, it is rather difficult to prescribe a real physical meaning to eq. (33) in the framework of CED, and at least semi-classical approaches to the description of material media (see, e.g. [24]) should be applied.

**Appendix A. Calculation of total time derivative of field momentum, eq. (17):**

$$\frac{d(P_{EM})_y}{dt} = -\frac{1}{4\pi c}\int_{V_0} E_x\left(\frac{\partial \dot{A}_y}{\partial x} - \frac{\partial \dot{A}_x}{\partial y}\right)dxdydz$$

We evaluate the first integral in eq. (17), using eqs. (16) and (18a-b):

$$-\frac{1}{4\pi c}\int_{-Y_0}^{Y_0}\int_{-Z_0}^{Z_0}\int_{-X_0}^{X_0} E_x \frac{\partial \dot{A}_y}{\partial x} dxdzdy = \frac{E\omega m_0}{4\pi c}\int_{-Y_0}^{Y_0}\int_{-Z_0}^{Z_0}\left(\left(\frac{(x-x_0)}{\left((x-x_0)^2+y^2+z^2\right)^{3/2}}\right)_{-X_0}^{0} - \left(\frac{(x-x_0)}{\left((x-x_0)^2+y^2+z^2\right)^{3/2}}\right)_{0}^{X_0}\right)dzdy =$$

$$\frac{E\omega m_0}{4\pi c}\int_{-Y_0}^{Y_0}\int_{-Z_0}^{Z_0}\left(-\frac{2x_0}{\left((x_0)^2+y^2+z^2\right)^{3/2}}+\frac{(X_0+x_0)}{\left((X_0+x_0)^2+y^2+z^2\right)^{3/2}}-\frac{(X_0-x_0)}{\left((X_0-x_0)^2+y^2+z^2\right)^{3/2}}\right)dzdy$$
(A1)

Using the expressions

$$\int \frac{dx}{\left(a^2+x^2\right)^{3/2}} = \frac{x}{a^2\left(a^2+x^2\right)^{1/2}}+const,$$  (A2)

$$\int \frac{dx}{\left(a^2+x^2\right)^{1/2}\left(b^2+x^2\right)} = \frac{1}{b\sqrt{a^2-b^2}}\arctan\frac{x\sqrt{a^2-b^2}}{b\left(\sqrt{a^2+x^2}\right)}+const,$$ (A3)

and adopting hereinafter $X_0=Y_0=Z_0$ for simplicity, we further obtain:

$$-\frac{E\omega m_0}{4\pi c}\int_{-Y_0}^{Y_0}\int_{-Z_0}^{Z_0}\frac{2x_0}{\left(x_0^2+y^2+z^2\right)^{3/2}}dzdy = -\frac{E\omega m_0 Z_0 x_0}{\pi c}\int_{-Y_0}^{Y_0}\frac{dy}{\left(x_0^2+y^2+Z_0^2\right)^{1/2}\left(x_0^2+y^2\right)} =$$

$$-\frac{E\omega m_0}{\pi c}\arctan\frac{yZ_0}{x_0\left(x_0^2+y^2+Z_0^2\right)^{1/2}}\bigg|_{-Y_0}^{Y_0} = -\frac{E\omega m_0}{c};$$
(A4a)

$$\frac{E\omega m_0}{4\pi c}\int_{-Y_0}^{Y_0}\int_{-Z_0}^{Z_0}\frac{(X_0+x_0)}{\left((X_0+x_0)^2+y^2+z^2\right)^{3/2}}dzdy = \frac{E\omega m_0 Z_0}{2\pi c}\int_{-Y_0}^{Y_0}\frac{(X_0+x_0)}{\left((X_0+x_0)^2+y^2+Z_0^2\right)^{1/2}\left((X_0+x_0)^2+y^2\right)}dy =$$

$$\frac{E\omega m_0}{\pi c}\arctan\frac{yZ_0}{(X_0+x_0)\left((X_0+x_0)^2+y^2+Z_0^2\right)^{1/2}}\bigg|_{-Y_0}^{Y_0} = \frac{2E\omega m_0}{\pi c}\arctan\frac{1}{\sqrt{3}} = \frac{E\omega m_0}{3c};$$
(A4b)

$$-\frac{E\omega m_0}{4\pi c}\int_{-Y_0}^{Y_0}\int_{-Z_0}^{Z_0}\frac{(X_0-x_0)}{\left((X_0-x_0)^2+y^2+z^2\right)^{3/2}}dzdy = -\frac{E\omega m_0 Z_0}{2\pi c}\int_{-Y_0}^{Y_0}\frac{(X_0-x_0)}{\left((X_0-x_0)^2+y^2+Z_0^2\right)^{1/2}\left((X_0-x_0)^2+y^2\right)}dy =$$

$$-\frac{2E\omega m_0}{\pi c}\arctan\frac{1}{\sqrt{3}} = -\frac{E\omega m_0}{3c}.$$
(A4c)

Substituting eqs. (A4a-c) into eq. (A1), we arrive at the equality:

$$-\frac{1}{4\pi c}\int_{V_0} E_x \frac{\partial \dot{A}_y}{\partial x}dxdydz = -\frac{E\omega m_0}{c}.$$ (A5)

Next, we evaluate the second integral in eq. (17), again applying eqs. (16) and (18a-b):

$$\frac{1}{4\pi c}\int_{-X_0}^{X_0}\int_{-Z_0}^{Z_0}\int_{-Y_0}^{Y_0} E_x \frac{\partial \dot{A}_x}{\partial y}dydxdz = -\frac{Y_0 \omega m_0}{2\pi c}\int_{-X_0}^{X_0}\int_{-Z_0}^{Z_0} E_x\left(\frac{dzdx}{\left((x-x_0)^2+Y_0^2+z^2\right)^{3/2}}\right) =$$

$$-\frac{Y_0 Z_0 \omega m_0}{\pi c}\int_{-X_0}^{X_0} E_x\left(\frac{dx}{\left((x-x_0)^2+Y_0^2+Z_0^2\right)^{1/2}\left((x-x_0)^2+Y_0^2\right)}\right) = -\frac{\omega m_0 E}{\pi c}\left(\arctan\frac{(x-x_0)Z_0}{Y_0\left((x-x_0)^2+Y_0^2+Z_0^2\right)^{1/2}}\right)_{-X_0}^{0} -$$ (A6)

$$\frac{\omega m_0 E}{\pi c}\left(\arctan\frac{(x-x_0)Z_0}{Y_0\left((x-x_0)^2+Y_0^2+Z_0^2\right)^{1/2}}\right)_{0}^{X_0} = \frac{\omega m_0 E}{\pi c}\arctan\frac{2x_0 Z_0}{Y_0\left(x_0^2+Y_0^2+Z_0^2\right)^{1/2}} \to 0$$

at $x_0 \ll Y_0, Z_0$.

Summing up eqs. (A5) and (A6), we obtain the $y$-component of time derivative of field momentum:

$$d(P_{EM})_y/dt = -\frac{1}{4\pi c}\int_{V_0} E_x\left(\frac{\partial \dot{A}_y}{\partial x} - \frac{\partial \dot{A}_x}{\partial y}\right)dxdydz = -\frac{E\omega m_0}{c}$$ (A7)

## Appendix B. Calculation of total time derivative of field angular momentum, eq. (28)

$$\left(\frac{d\theta_{EM}}{dt}\right)_z = -\frac{1}{4\pi c}\int_V xE_x\left(\frac{\partial \dot{A}_y}{\partial x} - \frac{\partial \dot{A}_x}{\partial y}\right)dV$$

First of all, we present this integral in the form

$$\left(\frac{d\theta_{EM}}{dt}\right)_z = \frac{1}{4\pi c}\int_V (x - x_0 + x_0)E_x\left(\frac{\partial \dot{A}_x}{\partial y} - \frac{\partial \dot{A}_y}{\partial x}\right)dV = \frac{1}{4\pi c}\int_V (x - x_0)E_x\left(\frac{\partial \dot{A}_x}{\partial y} - \frac{\partial \dot{A}_y}{\partial x}\right)dV + x_0\frac{d(P_{EM})_y}{dt} \quad (B1)$$

Next, we evaluate the first integral in eq. (B1), using eqs. (16) and (18a):

$$\frac{1}{4\pi c}\int_{-X_0}^{X_0}\int_{-Z_0}^{Z_0}\int_{-Y_0}^{Y_0}(x-x_0)E_x\frac{\partial \dot{A}_x}{\partial y}dydzdx = -\frac{Y_0\omega m_0}{2\pi c}\int_{-X_0}^{X_0}\int_{-Z_0}^{Z_0}E_x\left(\frac{(x-x_0)dzdx}{\left((x-x_0)^2 + Y_0^2 + z^2\right)^{3/2}}\right) =$$

$$-\frac{Y_0Z_0\omega m_0}{\pi c}\int_{-X_0}^{X_0}E_x\left(\frac{(x-x_0)dx}{\left((x-x_0)^2 + Y_0^2 + Z_0^2\right)^{1/2}\left((x-x_0)^2 + Y_0^2\right)}\right) = \quad \text{(B1-1)}$$

$$\frac{Y_0Z_0\omega m_0 E}{\pi c}\left[\left(\frac{1}{Z_0}\text{arctanh}\frac{\left((x-x_0)^2 + Y_0^2 + Z_0^2\right)^{1/2}}{Z_0}\right)_{-X_0}^{0} - \left(\left(\frac{1}{Z_0}\text{arctanh}\frac{\left((x-x_0)^2 + Y_0^2 + Z_0^2\right)^{1/2}}{Z_0}\right)\right)_{0}^{X_0}\right] = \frac{2Y_0\omega m_0 E}{\pi c}\text{arctanh}\sqrt{2} = \frac{Y_0\omega m_0 E}{\pi c}\ln\frac{\sqrt{2}+1}{\sqrt{2}-1},$$

where we have used the expressions $\int\frac{xdx}{(a^2+x^2)^{1/2}(b^2+x^2)} = -\frac{1}{\sqrt{a^2-b^2}}\text{arctanh}\frac{x\sqrt{a^2+x^2}}{\left(\sqrt{a^2-b^2}\right)} + const$, and

$\text{arctanh}(x) = \frac{1}{2}\ln\frac{x+1}{x-1}$.

Now we calculate the second integral in eq. (B1), using eqs. (16) and (18b):

$$-\frac{1}{4\pi c}\int_{V_0}(x-x_0)E_x\frac{\partial \dot{A}_y}{\partial x}dV = -\frac{1}{4\pi c}\int_{V_0}E_x\left(\frac{\partial\left((x-x_0)\dot{A}_y\right)}{\partial x} - \dot{A}_y\right)dV = -\frac{1}{4\pi c}\int_{V_0}E_x\frac{\partial\left((x-x_0)\dot{A}_y\right)}{\partial x}dV + \frac{1}{4\pi c}\int_{V_0}E_x\dot{A}_ydV,$$

(B1-2)

and evaluate separately the terms $-\frac{1}{4\pi c}\int_{V_0}E_x\frac{\partial\left((x-x_0)\dot{A}_y\right)}{\partial x}dV$ and $\frac{1}{4\pi c}\int_{V_0}E_x\dot{A}_ydV$. First we calculate

$$-\frac{1}{4\pi c}\int_{-Z_0}^{Z_0}\int_{-Y_0}^{Y_0}\int_{-X_0}^{X_0}E_x\frac{\partial\left((x-x_0)\dot{A}_y\right)}{\partial x}dxdydz =$$

$$\frac{E\omega m_0}{4\pi c}\int_{-Z_0}^{Z_0}\int_{-Y_0}^{Y_0}\left[\left(\frac{(x-x_0)^2}{\left((x-x_0)^2+y^2+z^2\right)^{3/2}}\right)_{-X_0}^{0} - \left(\frac{(x-x_0)^2}{\left((x-x_0)^2+y^2+z^2\right)^{3/2}}\right)_{0}^{X_0}\right]dydz = \quad \text{(B2-1)}$$

$$\frac{E\omega m_0}{4\pi c}\int_{-Z_0}^{Z_0}\int_{-Y_0}^{Y_0}\left(\frac{2x_0^2}{\left(x_0^2+y^2+z^2\right)^{3/2}} - \frac{(X_0+x_0)^2}{\left((X_0+x_0)^2+y^2+z^2\right)^{3/2}} - \frac{(X_0-x_0)^2}{\left((X_0-x_0)^2+y^2+z^2\right)^{3/2}}\right)dydz$$

Next, we subsequently evaluate the integrals in eq. (B2-1), using eqs. (A2), (A3) and (16):

$$\frac{E\omega m_0}{4\pi c}\int_{-Z_0}^{Z_0}\int_{-Y_0}^{Y_0}\frac{2x_0^2 dydz}{\left(x_0^2+y^2+z^2\right)^{3/2}} = \frac{E\omega m_0 Y_0 x_0^2}{\pi c}\int_{-Z_0}^{Z_0}\frac{dz}{\left(x_0^2+Y_0^2+z^2\right)^{1/2}\left(x_0^2+z^2\right)} =$$

(B2-1a)

$$\frac{E\omega m_0 x_0}{\pi c}\left(\arctan\frac{zY_0}{x_0\left(x_0^2+Y_0^2+z^2\right)^{1/2}}\right)_{-Z_0}^{Z_0} = \frac{E\omega m_0 x_0}{c};$$

$$-\frac{E\omega m_0(X_0+x_0)^2}{4\pi c}\int_{-Z_0}^{Z_0}\int_{-Y_0}^{Y_0}\frac{1}{\left((X_0+x_0)^2+y^2+z^2\right)^{3/2}}dydz=$$

$$-\frac{E\omega Y_0 m_0(X_0+x_0)^2}{2\pi c}\int_{-Y_0}^{Y_0}\frac{1}{\left((X_0+x_0)^2+Y_0^2+z^2\right)^{1/2}}\frac{1}{\left((X_0+x_0)^2+z^2\right)^{1/2}}dz= \quad\text{(B2-1b)}$$

$$-\frac{E\omega m_0(X_0+x_0)}{2\pi c}\arctan\frac{zY_0}{(X_0+x_0)\left((X_0+x_0)^2+Y_0^2+z^2\right)^{1/2}}=-\frac{E\omega m_0(X_0+x_0)}{\pi c}\arctan\frac{1}{\sqrt{3}}=-\frac{E\omega m_0(X_0+x_0)}{6c};$$

$$-\frac{E\omega m_0(X_0-x_0)^2}{4\pi c}\int_{-Z_0}^{Z_0}\int_{-Y_0}^{Y_0}\frac{1}{\left((X_0-x_0)^2+y^2+z^2\right)^{3/2}}dydz=$$

$$-\frac{E\omega Y_0 m_0(X_0-x_0)^2}{2\pi c}\int_{-Z_0}^{Z_0}\frac{1}{\left((X_0-x_0)^2+Y_0^2+z^2\right)^{1/2}}\frac{1}{\left((X_0+x_0)^2+z^2\right)^{1/2}}dz= \quad\text{(B2-1c)}$$

$$-\frac{E\omega m_0(X_0-x_0)}{2\pi c}\arctan\frac{zY_0}{(X_0-x_0)\left((X_0+x_0)^2+Y_0^2+z^2\right)^{1/2}}=$$

$$-\frac{E\omega m_0(X_0-x_0)}{\pi c}\arctan\frac{1}{\sqrt{3}}=-\frac{E\omega m_0(X_0-x_0)}{6c}.$$

Summing up eqs. (B2-1a), (B2-1b) and (B2-1c), we obtain:

$$-\frac{1}{4\pi c}\int_V E_x\frac{\partial\left((x-x_0)\dot A_y\right)}{\partial x}dydzdx=\frac{E\omega m_0 x_0}{c}-\frac{E\omega m_0(X_0+x_0)}{6c}-\frac{E\omega m_0(X_0-x_0)}{6c}=\frac{E\omega m_0 x_0}{c}-\frac{E\omega m_0 X_0}{3c} \quad\text{(B3)}$$

Next, we calculate the second integral in eq. (B1-2):

$$\frac{1}{4\pi c}\int_V E_x\dot A_y dydzdx=-\frac{E\omega m_0}{4\pi c}\int_{-Z_0}^{+Z_0}\int_{-Y_0}^{+Y_0}\left(\int_{-\infty}^{0}\frac{x-x_0}{\left((x-x_0)^2+y^2+z^2\right)^{3/2}}dx-\int_{0}^{\infty}\frac{x-x_0}{\left((x-x_0)^2+y^2+z^2\right)^{3/2}}dx\right)dydz=$$

$$=\frac{E\omega m_0}{4\pi c}\int_{-Z_0}^{+Z_0}\int_{-Y_0}^{+Y_0}\left(\left(\frac{1}{\left((x-x_0)^2+y^2+z^2\right)^{1/2}}\right)_{-X_0}^{0}-\left(\frac{1}{\left((x-x_0)^2+y^2+z^2\right)^{1/2}}\right)_{0}^{X_0}\right)dydz= \quad\text{(B4)}$$

$$\frac{E\omega m_0}{4\pi c}\int_{-Z_0}^{+Z_0}\int_{-Y_0}^{+Y_0}\left(\frac{2}{\left(x_0^2+y^2+z^2\right)^{1/2}}-\frac{1}{\left((X_0+x_0)^2+y^2+z^2\right)^{1/2}}-\frac{1}{\left((X_0-x_0)^2+y^2+z^2\right)^{1/2}}\right)dydz,$$

where we have used the expression $\int\frac{xdx}{\left(a^2+x^2\right)^{3/2}}dx=-\frac{1}{\left(a^2+x^2\right)^{1/2}}+const$.

Calculate the first integral in eq. (B4):

$$\frac{E\omega m_0}{4\pi c}\int_{-Z_0}^{+Z_0}\int_{-Y_0}^{+Y_0}\frac{2}{\left(x_0^2+y^2+z^2\right)^{1/2}}dydz=\frac{E\omega m_0}{2\pi c}\int_{-Z_0}^{+Z_0}\ln\left(\frac{\left(x_0^2+Y_0^2+z^2\right)^{1/2}+Y_0}{\left(x_0^2+Y_0^2+z^2\right)^{1/2}-Y_0}\right)dz=$$

$$\frac{E\omega m_0}{2\pi c}\left(-2x_0\arctan\frac{Y_0 z}{x_0\left(x_0^2+Y_0^2+z^2\right)^{1/2}}+z\ln\left(\frac{\left(x_0^2+Y_0^2+z^2\right)^{1/2}+Y_0}{\left(x_0^2+Y_0^2+z^2\right)^{1/2}-Y_0}\right)-Y_0\ln\left(\frac{\left(x_0^2+Y_0^2+z^2\right)^{1/2}+z}{\left(x_0^2+Y_0^2+z^2\right)^{1/2}-z}\right)\right)\Bigg|_{-Z_0}^{Z_0} \quad\text{(B4-1)}$$

$$=-\frac{E\omega m_0 x_0}{c},$$

where we have used the expression:

$$\int\frac{dx}{\left(a^2+x^2\right)^{1/2}}dx=\ln\left(\left(a^2+x^2\right)^{1/2}+x\right)+const,$$

$$\int\ln\left(\left(a^2+x^2\right)^{1/2}+b\right)=-\sqrt{a^2+b^2}\arctan\frac{bx}{\sqrt{a^2+b^2}\sqrt{a^2+x^2}}+\sqrt{a^2+b^2}\arctan\frac{x}{\sqrt{a^2-b^2}}+$$

$$x\ln\left(\left(a^2+x^2\right)^{1/2}+b\right)+b\ln\left(\left(a^2+x^2\right)^{1/2}+x\right)-x+const,$$

$$\int\ln\left(\left(a^2+x^2\right)^{1/2}-b\right)=\sqrt{a^2+b^2}\arctan\frac{bx}{\sqrt{a^2+b^2}\sqrt{a^2+x^2}}+\sqrt{a^2+b^2}\arctan\frac{x}{\sqrt{a^2-b^2}}+$$

$$x\ln\left(\left(a^2+x^2\right)^{1/2}-b\right)-b\ln\left(\left(a^2+x^2\right)^{1/2}+x\right)-x+const.$$

By analogy we calculate the second integral in eq. (B4):

$$-\frac{E\omega m_0}{4\pi c}\int_{-Z_0}^{+Z_0}\int_{-Y_0}^{+Y_0}\left(\frac{1}{\left((X_0+x_0)^2+y^2+z^2\right)^{1/2}}\right)dydz=$$

$$-\frac{E\omega m_0}{4\pi c}\left(-2(X_0+x_0)\arctan\frac{Y_0 z}{(X_0+x_0)\left((X_0+x_0)^2+Y_0^2+z^2\right)^{1/2}}\right)_{-Z_0}^{Z_0}= \quad \text{(B4-2)}$$

$$\frac{E\omega m_0(X_0+x_0)}{\pi c}\arctan\frac{1}{\sqrt{3}}=\frac{E\omega m_0(X_0+x_0)}{6c},$$

as well as the third integral in eq. (B4):

$$-\frac{E\omega m_0}{4\pi c}\int_{-Z_0}^{+Z_0}\int_{-Y_0}^{+Y_0}\left(\frac{1}{\left((X_0-x_0)^2+y^2+z^2\right)^{1/2}}\right)dydz=$$

$$-\frac{E\omega m_0}{4\pi c}\left(-2(X_0-x_0)\arctan\frac{Y_0 z}{(X_0-x_0)\left((X_0-x_0)^2+Y_0^2+z^2\right)^{1/2}}\right)_{-Z_0}^{Z_0} \quad \text{(B4-3)}$$

$$=\frac{E\omega m_0(X_0-x_0)}{\pi c}\arctan\frac{1}{\sqrt{3}}=\frac{E\omega m_0(X_0-x_0)}{6c}.$$

Hence, the sum of (B4-1), (B4-2) and (B4-3) yields:

$$\frac{1}{4\pi c}\int_V E_x\dot{A}_y dydzdx=-\frac{E\omega m_0 x_0}{c}+\frac{E\omega m_0 X_0}{3c} \quad \text{(B5)}$$

Substituting eqs. (B5), (B3) into eq. (B1-2), we obtain:

$$-\frac{1}{4\pi c}\int_{V_0}(x-x_0)E_x\frac{\partial \dot{A}_y}{\partial x}dV=-\frac{1}{4\pi c}\int_{V_0}E_x\frac{\partial\left((x-x_0)\dot{A}_y\right)}{\partial x}dV+\frac{1}{4\pi c}\int_{V_0}E_x\dot{A}_y dV=$$

$$\frac{E\omega m_0 x_0}{c}-\frac{E\omega m_0 X_0}{3c}-\frac{E\omega m_0 x_0}{c}+\frac{E\omega m_0 X_0}{3c}=0. \quad \text{(B6)}$$

Further substitution of eqs. (B6), (B1-1) and (A7) into eq. (B1) yields:

$$\left(\frac{d\theta_{EM}}{dt}\right)_z=-\frac{1}{4\pi c}\int_V(x-x_0)E_x\left(\frac{\partial \dot{A}_x}{\partial y}-\frac{\partial \dot{A}_y}{\partial x}\right)dV+x_0 P_y=\frac{Y_0\omega m_0 E}{\pi c}\ln\frac{\sqrt{2}+1}{\sqrt{2}-1}-\frac{E\omega m_0 x_0}{c} \quad \text{(B7)}$$

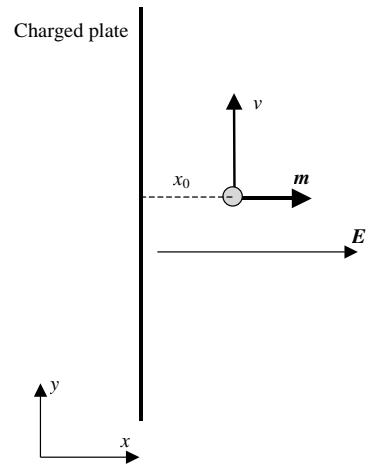

Fig. 1. Interaction of a moving magnetic dipole ***m*** with the resting homogeneously charged plate, lying in the plane *yz*. The velocity of dipole ***v*** is parallel to the axis *y*, and vector ***m*** is parallel to the electric field of the plate ***E***. At the initial time moment the dipole passes the axis *x* in the point $x_0$, and we want to check the implementation of conservation laws for the total momentum and total angular momentum of this isolated system.